\documentclass{stsci_report}

\usepackage{color}
\usepackage{graphicx}
\usepackage{longtable}
\usepackage{tabularray}
\usepackage{hyperref}
\usepackage{array}
\usepackage{indentfirst}
\usepackage{float}
\usepackage[title]{appendix}

\newcolumntype{R}{>{\raggedleft\arraybackslash}p{2cm}}
\newcolumntype{C}{>{\centering\arraybackslash}p{2cm}}

\copyrighttext{Copyright\copyright\ \the\year\ The Association of 
Universities for Research in Astronomy, Inc. All Rights Reserved.}

\presubtitle{ISR WFC3 2025-08}
\title{Pixel-Based Non-Linearity Correction for the WFC3 IR Detector}
\author{Sachindev S. Shenoy, Ky Huynh, Varun Bajaj, \& 
Jennifer Mack}
\date{November 4, 2025}
\begin{document}
\maketitle
\abstract{The current non-linearity correction for the Wide 
Field Camera 3 Infrared (WFC3/IR) channel is based on ground-based data 
acquired during WFC3's Thermal Vacuum 3 (TV3) testing campaign. In the 
current reference file, the correction coefficients derived for each 
pixel are averaged over each of the four detector quadrants. In this work, 
we compute a new pixel-based non-linearity correction using in-flight calibration 
observations with the internal tungsten lamp flats acquired 
between 2011 and 2013. We derive the new correction coefficients by 
fitting a third-order polynomial to the accumulated signal ``up-the-ramp" 
for each pixel. Approximately 2\% of IR detector pixels are flagged as bad, 
and a solution cannot be computed. For these, we use the quadrant averages 
of the new correction coefficients.  An accompanying report 
(\cite{huynh2025}) provides detailed testing results using both internal 
flats and external science targets acquired in a variety of observing modes. 
The report highlights improvements in photometry derived from ``\_flt.fits" 
data products calibrated using the new reference file, with the largest 
improvement for pixels with fluence levels approaching the full well limit 
of $\sim$80,000 $e^- $. A new NLINFILE reference file was delivered in 
October 2025 and will be used to reprocess all WFC3/IR imaging and grism 
 observations in the MAST archive.}

\clearpage

\section*{Introduction}
\label{sec:intro}
The Wide Field Camera 3 Infrared (WFC3/IR) channel  is a 1024 $\times$ 
1024 HgCdTe (Mercury Cadmium Telluride) detector built by Teledyne. 
HgCdTe is one of the most commonly used semiconductor materials to 
detect IR radiation due to the ability to tune the optical properties 
of the semiconductor to desired IR wavelengths by choosing the amount of 
Cd in the alloy (\cite{WFC3IHB2024}). IR radiation detection occurs when 
a photon impinges on the material, generating electron-hole pairs, where 
some of the electrons and holes recombine, and the remaining electrons 
(or holes, depending on the material) are kicked from the valence band to 
the conduction band. These electrons are then collected by the readout 
circuit and transformed into an electric signal. 
\bigbreak
Ideally, as the total number of incident photons increases, the number of 
electrons in the conduction band increases linearly. However, when the 
conduction band approaches saturation, the response of the electron 
deviates from linearity with respect to the incident photons. Due to 
defects in the material, it has been shown experimentally that the IR 
detector is inherently non-linear even at low numbers of incident photons. 
\bigbreak
To correct for the non-linearity effects, we first fit a linear response 
to low signal levels (reads) to get an ``ideal" signal for each pixel, 
which is then used to normalize the later reads. One can then fit a 
higher-order polynomial to the normalized signal and record the coefficients 
of the polynomial fit. These coefficients can then be used to correct the 
observed science data. The detailed description of this method is provided 
in the Model Fitting section below. As of 2025, the non-linearity correction 
coefficients in \texttt{calwf3} (WFC3 data processing pipeline) are derived 
using a similar method but are averaged over an entire quadrant of the 
detector, hereafter referred to as quad-based correction 
(\cite{hilbert2008}). In this document, we explore the improvement in the 
non-linearity correction achieved by deriving the correction coefficients 
on a per-pixel basis (hereafter referred to as the pixel-based correction), 
rather than the quad-based correction.

\section*{Data Reduction}
\label{sec:reduction}
The standard method to derive the IR non-linearity corrections is to 
observe flat field data with short and long exposure times, which provides 
one with both data in the ``linear" and ``non-linear" regimes, respectively. 
The WFC3/IR channel uses a MULTIACCUM mode in which the detector signal is 
read out non-destructively multiple times. There are 12 predefined sample 
sequences that accommodate a wide range of observing conditions. For this 
work, we use data from WFC3 in-flight calibration (CAL) programs designed 
to characterize the non-linearity by obtaining flat-field exposures with 
the internal tungsten lamps using the SPARS25 sample sequence (\href{https://hst-docs.stsci.edu/wfc3ihb/chapter-7-ir-imaging-with-wfc3/7-7-ir-exposure-and-readout#id-7.7IRExposureandReadout-7.7.37.7.3MULTIACCUMTimingSequences:FullArrayApertures}{Section 7.7.3 - 
WFC3 Instrument Handbook}). This sequence is ideal for characterizing the 
linearity because it samples (reads out) the detector with a uniform 
interval of $\sim$25 seconds. Each flat is read out 15 times, the maximum 
allowed for the WFC3/IR channel, allowing the majority of detector pixels 
to become saturated in the last few reads of the exposure. We refer to the 
set of samples (reads) for an individual exposure as a ``ramp" in the rest 
of this document. 
\bigbreak
We use a similar dataset that was used in a previous WFC3/IR non-linearity 
correction study  (\cite{hilbert2014}), focusing only on the SPARS25 
internal flats. All of the data are part of the non-linearity monitoring 
WFC3/CAL programs (proposal IDs 12352, 12696, \& 13079), which consist of 
both internal flats and external star cluster data. Internal flats are 
first acquired in the F126N filter to warm up the lamp and achieve a 
stable flux. Flats are then acquired in the F127M filter to characterize 
the non-linearity. Table~\ref{tab:dataidflat} in Appendix A lists the 
proposal ID, filter, sample sequence, exposure times, and rootnames of 
all F127M flat field and dark observations used. Internal flats from the 
first non-linearity CAL program, 11931, did not allow for sufficient lamp 
warm-up time and were therefore excluded from both studies.
\bigbreak
Previous calibration observations (\cite{long2011,long2018}) have shown 
that WFC3/IR suffers from persistence issues, and one needs to plan for 
persistence mitigation during observations (\cite{long2010}). Dark frames 
are obtained before and after the flat observations to monitor persistence. 
This was fortuitous, allowing us to use the zeroth read (the detector 
readout at 2.932 sec after the start of the dark observation) as the bias 
level in our data reduction. In the normal WFC3/IR data reduction pipeline, 
\texttt{calwf3}, the reference pixels located at the periphery of the 
detector are used to compute a $\sigma$-clipped mean signal, which is used
as a bias level. While this is a good estimation of the detector bias level, 
it does not represent the pixel-to-pixel variations in the bias signal, 
which we must account for when deriving a pixel-based non-linearity 
correction. We thus elected to use the zeroth read of the dark observations 
as the bias, rather than the $\sigma$-clipped mean calculated from the 
reference pixels.
\bigbreak
In addition to the lamp flats, these WFC3/CAL programs also included 
external observations of the star cluster 47 Tucanae, consisting of both 
short and long exposures in the F160W filter.  These observations were 
designed to test the non-linearity solution derived from lamp flats and 
are discussed in an accompanying report (\cite{huynh2025}), which compares 
photometry derived with the new pixel-based non-linearity correction with 
the current quad-based correction. 
\bigbreak
The general procedure for deriving the non-linearity correction coefficients 
is listed below, and the details are described in the following sections:
\begin{itemize}
\item Download the data (RAW, and IMA data products) from the MAST archive. 
A total of 23 F127M flat field exposures and 23 corresponding dark frames 
from the three aforementioned CAL programs were used to derive the 
pixel-based non-linearity correction (see Table~\ref{tab:dataidflat}).
\item Group the data into individual visits, which include the RAW and IMA 
data products for the internal flat and the corresponding dark.
\item Examine the IMA data products to identify pixels flagged in the DQ 
array and to estimate the percentage of detector pixels that may fail 
during the fitting step. Pixels with a DQ flag value = 8192, corresponding 
to cosmic ray (CR) hits, were specifically inspected. These 
can produce discontinuities or jumps in the ramp signal, which can 
adversely affect the fitting process. 
\item Perform a bias subtraction using the zeroth read of the preceding 
dark observation as a bias frame.
\item Mean combine each of the ramps for all the flat observations to 
generate a master flat. 
\item Flag all pixels that are either dead, saturated in early reads, or hard 
saturated (defined as pixels exceeding 25\% non-linearity).
\item Fit a linear model to the first three reads to estimate an ideal 
response. For more details, see the Model Fitting section, Equation ~\ref{eq:ideal}, and Figure~\ref{fig:linfit}.
\item Normalize the master flat with the ideal response to derive a 
fractional linearity. 
\item Fit a polynomial equation to the fractional linearity to derive the 
non-linearity correction coefficients.
\item For bad pixels, which are masked and have a NaN value, use the 
quadrant-based averages of the new coefficients as the correction 
coefficients. 
\item Generate the non-linearity correction reference file with an 
appropriate format that can be used with the WFC3 data reduction pipeline, 
\texttt{calwf3}. For details on the file format, see 
Table~\ref{tab:data_format}.
\end{itemize}

\subsection*{Data Cleaning}
\label{sec:cleaning}
All of the RAW and calibrated IMA files for both the internal flat 
observations and their associated darks were downloaded using the 
\texttt{astroquery} Application Programming Interface (API). Inspection 
of the DQ flags in the IMA files showed that, on average, 1.95 $\pm$ 
0.01 \% of the pixels in a given read are flagged for various reasons 
such as dead pixels (DQ=4), hot pixels (DQ=16), or unstable pixels 
(DQ=32). See Table 2.7 in Section 2.2.3 in \cite{WFC3DHB2024} for a 
complete list of DQ flags. Given the large number of input observations 
available for generating the master flat, we possess sufficient statistics 
to detect and exclude outliers (bad pixels) during the merging process. We 
therefore did not flag any bad pixels at this stage; however, we did 
investigate those pixels identified as cosmic ray hits.
\bigbreak
Cosmic ray (CR) impacts are flagged in the IMA file (DQ=8192) during the 
``up-the-ramp" fitting step in $\texttt{calwf3}$, which looks for jumps in 
the rate signal that deviates by more than 4-sigma from the expected linear 
response (see Section 3.3.10 of the WFC3 Data Handbook, \cite{WFC3DHB2024}). 
We find that SPARS25 flats calibrated using the 2008 quadrant-based 
reference file 
have overly aggressive CR flagging, setting the DQ flag for a large fraction 
of detector pixels. Once a CR flag is set in a given read of the IMA file, 
all subsequent reads for that pixel are flagged. Furthermore, the majority 
of these pixels had additional DQ flags set, complicating the isolation of 
pure CR impacts. This issue precludes masking pixels flagged as CR hits, as 
this would exclude a substantial number ($\approx$ 30\%) of pixels from 
the linearity fitting. 
Further investigation into the pipeline’s CR flagging settings and the 
appropriate rejection threshold for different IR sample sequences is beyond 
the scope of this document.

\subsection*{Bias subtraction} 
\label{sec:biasub}
The WFC3/IR channel exhibits a bias signal arising from the DC offset 
applied to the detector, fixed pattern noise, and readout noise. A nominal 
step in IR data processing is to subtract a bias frame from each input data 
frame. For WFC3/IR, we must subtract this bias signal from each of the ramp 
reads. Therefore, a dedicated bias observation concurrent with the flat 
observation is essential for obtaining an accurately bias-subtracted flat. 
Since all calibration programs included a dark frame before and after the 
flat observations, we utilized the zeroth read of the preceding dark frame 
as our bias signal. We did not use the zeroth read of the flat observation 
as a bias signal, despite it having the same readout time ($\sim3\ s$) as 
the dark, because it includes additional signal from the internal lamp. 
Although a master bias generated from all dark observations would increase 
the signal-to-noise ratio, its noise profile is sufficiently different from 
that of the individual flat, which would introduce a non-optimal noise 
profile during subtraction.
\begin{figure}[h!]
    \centering
    \includegraphics[width=1.0\linewidth]{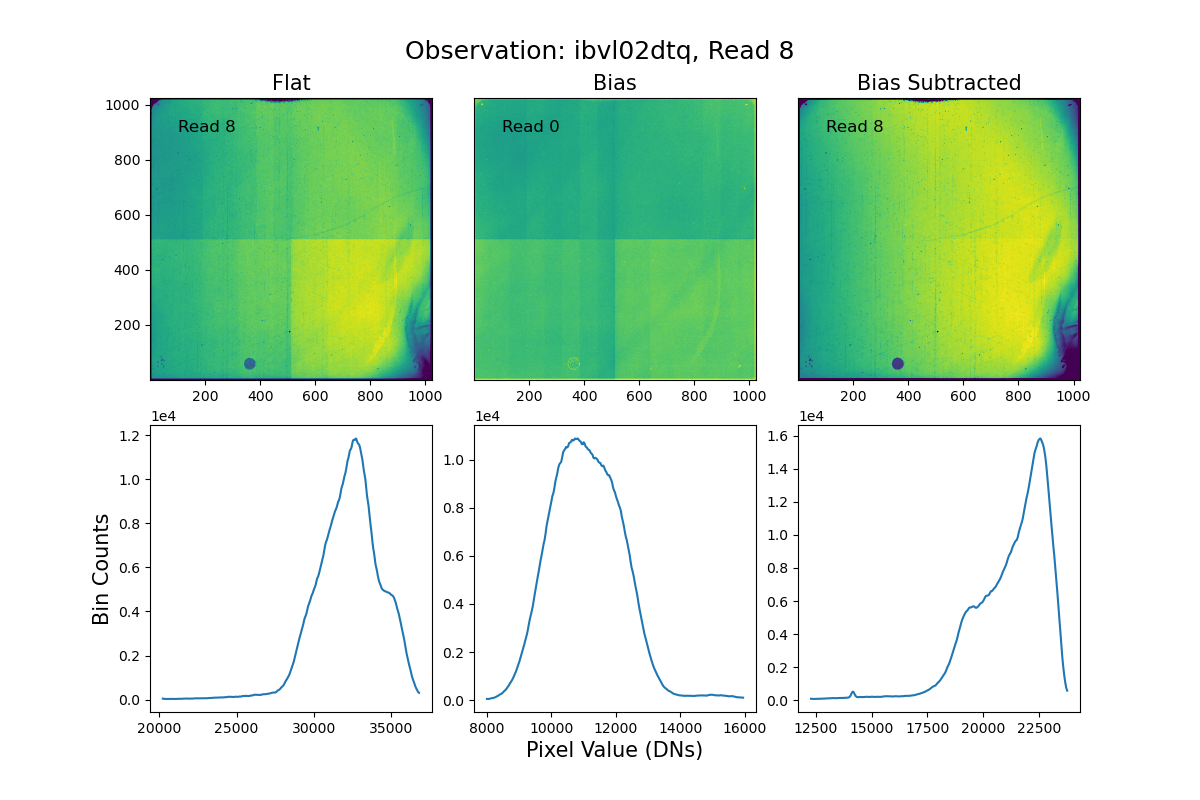}
    \vspace*{-0.5in}
    \caption{\small Example of bias-subtracted data. The first column shows
    the flat SCI array, the middle column is the bias frame (zeroth read of 
    a RAW dark file), and the last column is the bias-subtracted flat. The 
    second row is a histogram of pixel values for each respective frame. 
    Images in the top row are displayed in a linear scale with zscale 
    interval.} 
    \label{fig:pbsf_dat}
\end{figure}

Figure~\ref{fig:pbsf_dat} shows an example of bias subtraction, where the 
left panel shows read 8 of the RAW flat data, the middle panel shows read 0 
of the dark data used as a bias signal, and the right panel shows read 8 of 
the bias-subtracted data. Analysis of the bias-subtracted flats showed 
that overall, this process produced a nominal flat. However, in each of 
the individual observations, a small subset of pixels had very low pixel 
values, resulting either from a non-responsive pixel in the flat or an 
abnormally high value in the bias. These pixels would get flagged by 
the outlier rejection when generating the master flat, but will need 
additional scrutiny during the fitting process.

\subsection*{Generate Master Flat} 
\label{sec:master}
Ideally, one can fit a polynomial model to the individual bias-subtracted 
flat, but due to the higher noise in the individual observations, the fits 
are not well constrained, which results in derived correction coefficients 
with higher uncertainties. Assuming that the read noise and non-linear 
response do not vary from image to image, we combine the bias-subtracted 
flats to generate a master flat before fitting a polynomial model. Each of 
the ramps in the bias-subtracted flats were combined using a 3-$\sigma$ 
clipped mean with associated uncertainties propagated. In the previous 
steps, we have not masked any bad pixels, as the outlier rejection will 
exclude most of these bad pixels. However, inspection of the master flat 
showed that three kinds of pixel responses needed to be flagged and excluded 
from the fitting. These three non-optimal responses are:
\begin{description}
\item[Early saturated pixels:] Pixels that reach the maximum fluence level 
(saturation) within the first two reads. In the master flat, we detected 
about 100 pixels that reached maximum fluence values in the first read and 
about 50 pixels in the second read. Since the saturation was reached in 
the early reads, we flagged all the reads on this set and excluded them 
from fitting.
\item[Dead pixels:] Pixels whose fluence values were always below 100 
counts for all the reads. We found about a dozen pixels that showed this 
behavior.  
\item[Hard saturated pixels:] Pixels that reached the well depth limit 
before the end of the entire ramp sequence. Since not all of the pixels 
in a given detector are uniform, they tend to reach their well depth at 
different fluence levels. We detected these pixels by fitting a linear 
response to the first three reads and then estimating the percent 
difference for each pixel up the ramp. We flagged any pixel that was 25\% 
lower than the linear model to indicate that these pixels were reaching 
their well depth limits. For most of the non-linearity coefficient 
calculations in the IR detector, it is nominal to set a pixel as saturated 
when it is 5\% lower than the linear model. So our limit of 25\% is well 
beyond this nominal value and will probe the highly non-linear regime as 
the pixel approaches the well depth limit. 
\end{description}

\begin{figure}[H]
    \centering
    \includegraphics[width=0.8\linewidth]{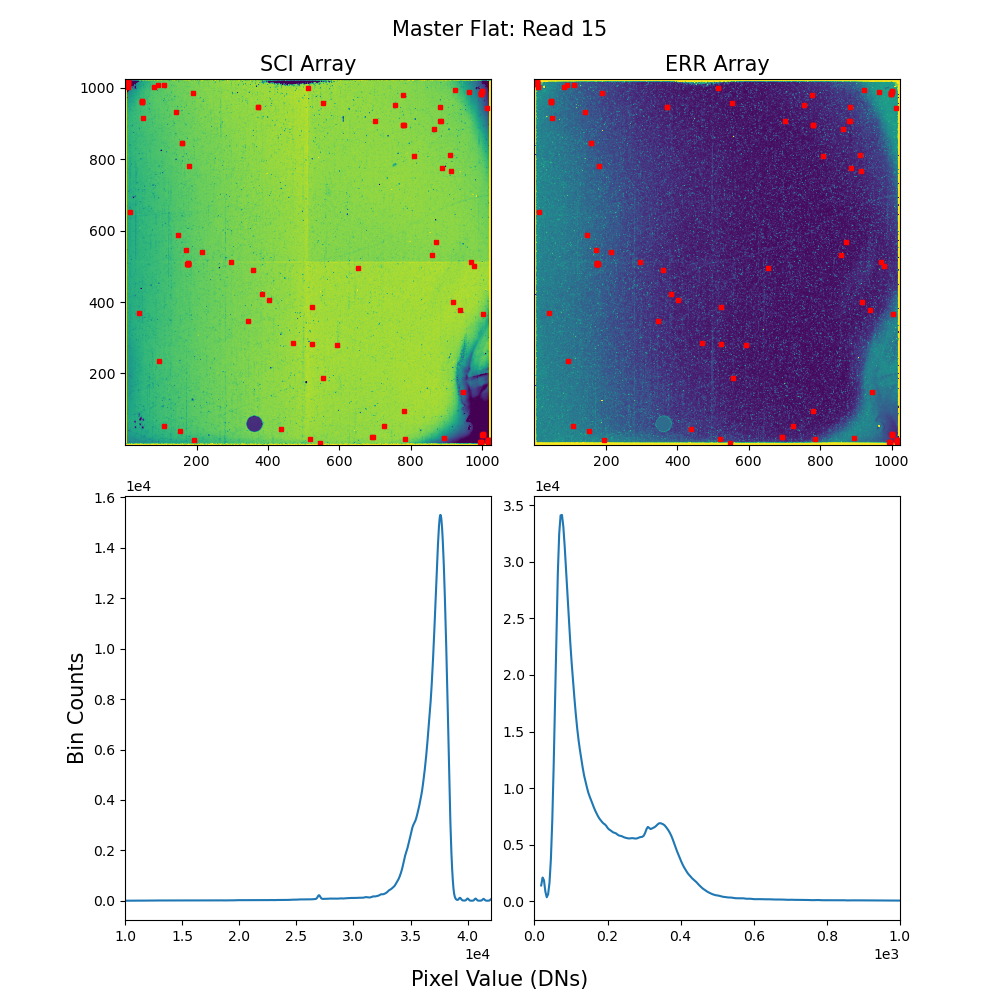}
    \vspace*{-0.2in}
    \caption{\small The top row displays the cumulative read of the SCI 
    and ERR arrays from the master flat. The red squares are early 
    saturated pixels, dead pixels, and hard saturated pixels. The bottom 
    row is the histogram of pixel values in the SCI and ERR arrays, 
    respectively. The images in the top row are
    displayed in a linear scale with zscale interval.}
    \label{fig:mast_flat}
\end{figure}

We save the indices of these flagged pixels to be used later in the 
reduction process. Figure~\ref{fig:mast_flat} (top panels) shows the SCI 
and ERR arrays of the last read of the master flat file, along with the 
location of flagged pixels. The bottom panel in the figure shows a 
histogram of the pixel value (DN) in the SCI and ERR arrays, respectively. 

\subsection*{Model Fitting}
\label{sec:fitting}
\begin{figure}[h!]
    \centering
    \includegraphics[width=0.7\linewidth]{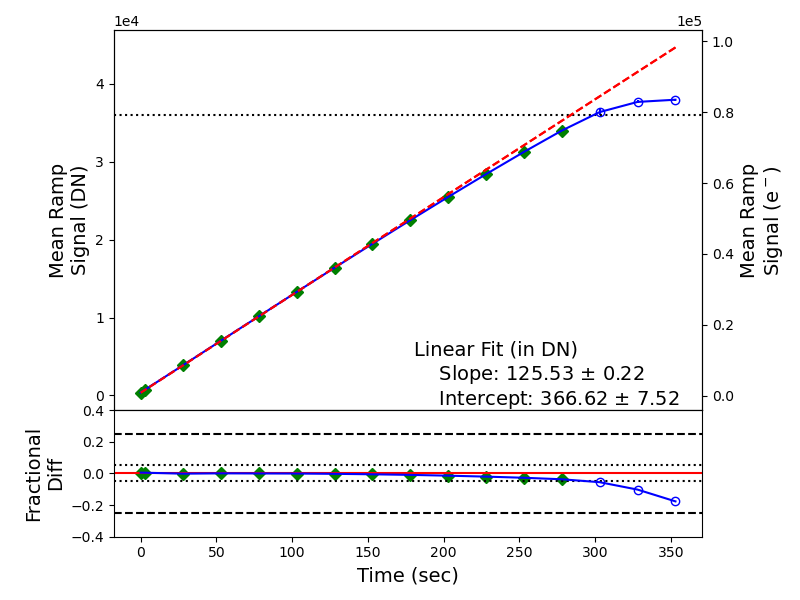}
    \vspace*{-0.2in}
    \caption{WFC3/IR non-linear response of the mean signal (DN) in the 
    master flat as a function of sample time (sec). Top Panel: Shows the 
    mean signal in the master flat for a sample pixel (indices 465, 987). 
    The green diamonds represent the mean signal values. The red dashed 
    line is the linear fit derived from the first four reads. The 
    horizontal black dotted line indicates the pixel's saturation limit. 
    Blue open circles highlight mean signal values exceeding this saturation 
    limit. 
    Bottom Panel: Displays the fractional difference between the mean 
    signal and the linear fit shown in the top panel. The dotted and 
    dashed lines indicate the 5\% and 25\% difference thresholds, 
    respectively.}
    \label{fig:linfit}
\end{figure}

After merging and cleaning the data, we start the determination of the 
non-linear coefficients  by fitting an ideal response (see 
Figure~\ref{fig:linfit}) and using it to calculate the fractional 
linearity.  Since we already fit a linear function to early reads of the 
ramp in order to flag pixels that reach their well depth limit (see the 
``Generate Master Flat" section), we can use the same fit results to get 
an ideal signal across all reads and use it to determine fractional 
linearity. The ``ideal" signal is given by:
\begin{equation}
\label{eq:ideal}
y_{ij\; ideal} = m_{ij} * t_{ij} + c_{ij}
\end{equation}
where $y_{ij\; ideal}$ is the measured signal for pixel index ($i$, $j$) at 
time $t$, $m_{ij}$ and $c_{ij}$ are the slope and intercept of the linear 
model, respectively. \\
Note that the use of this ideal response equation relies on two primary 
assumptions.
\bigbreak
The first assumption is that the first three (or more) reads exhibit a 
linear response, a premise that is often invalid. Previous studies 
(\cite{robberto2010,robberto2011}) have suggested that no fluence regime 
shows purely linear pixel response, with non-linear behavior occurring 
even at the lowest fluence levels. Recently, test using data from SPARS10 
flats show that non-linearity effects at low fluence levels are on the 
order of $1\%$ (see \cite{huynh2025}, Figure 4). However, due to the sparse 
sampling sequence of SPARS25 at low fluence levels, we are unable to 
determine the higher-order coefficients and must, therefore, assume a linear 
response in this region.
\bigbreak
We investigated the possibility of merging our data with flat observations 
taken using a different sample sequence with shorter read intervals. This, 
in principle, will yield additional data points in the low fluence region 
and will help in determining higher-order terms in the polynomial fit.  
Though, we found that the merged dataset resulted in an offset between 
data from the two sample sequences. Since the bias levels for each pixel 
is different between the two sample sequences, merging the datasets is not 
trivial. As such, we chose to use only the one sample sequence (SPARS25) 
and maintain the assumption of a linear response in the low fluence region.
To further explore these issues we plan to get additional calibration 
observations using SPARS5 flats that will be acquired in Cycle 33 (program 
17957). 
\bigbreak
Our second assumption involves the use of an intercept term ($c_{i,j}$), 
despite the input data being bias-subtracted. This is essential because 
the bias subtraction is imperfect, and the intercept accounts for any 
residual offset. Allowing a non-zero intercept ensures that even pixels 
affected by large errors in the bias subtraction will yield reasonably 
accurate fit coefficients.
\bigbreak
Figure~\ref{fig:linfit} shows an example of the non-linear response for a 
pixel with indices (465, 987) in the master flat. The plot shows the mean 
ramp signal (in DN) as a function of time in the top panel and the fractional 
difference between the measured data and the linear fit in the bottom panel. 
The diamond data points in the top panel are the measured signal, the red 
dashed line is the linear fit to the first four reads (roughly for signal up 
to 10,000 DN), and the black dotted line is the saturation limit for this 
pixel where the measured signal deviates more than 5\% from the linear fit. 
In both panels, the green diamonds and blue open circles refer to data that 
is below and above the saturation limit (percent difference $<$ 5\%), 
respectively. (A map of the full well limit is provided in Figure \ref{fig:f_satmap} 
of Appendix C and highlights both pixel-to-pixel and 
spatial variations across the detector.)
\bigbreak
Next, we define the fractional linearity (Equation ~\ref{eq:frac}) as 
normalizing the observed signal by the ideal signal, which 
is identical to the  factional difference data shown in the bottom panel 
of Figure~\ref{fig:linfit}. As seen in the figure, this residual signal
will now consist of only the non-linear terms and can be used to fit a 
higher-order polynomial.\\
 \begin{equation}
 \label{eq:frac}
 F_{ij\; Lin} = \frac{y_{ij\; obs} }{y_{ij\; ideal}}
 \end{equation}\\
Lastly, we use a third-order polynomial to fit the fractional linearity 
and derive the non-linearity correction coefficients, as shown in 
Equation ~\ref{eq:polyfit}). \\
\begin{equation}
\label{eq:polyfit}
 F_{ij\; Lin}  =  1 - (\alpha_{ij} + \beta_{ij} * y_{ij\; obs} + 
 \gamma_{ij} * y_{ij\; obs}^2 + 
 \delta_{ij} * y_{ij\; obs}^3)
\end{equation}\\
where $\alpha_{ij}$, $\beta_{ij}$, $\gamma_{ij}$, and $\delta_{ij}$ are the 
fit coefficients and $y_{ij\; obs}$ is the observed signal in the ramp.
These derived coefficients are then used to correct the observed 
signal, a process implemented in the $\texttt{NLINCORR}$ step of 
$\texttt{calwf3}$ (\S~3.3.7 in \cite{WFC3DHB2024}).
\bigbreak
We now have a set of coefficients that correct the non-linear response 
for most pixels, with two exceptions: 
\begin{itemize}
    \item Non-nominal pixels that were masked after the master flat 
    derivation.
    \item Pixels where the curve fitting failed.
\end{itemize}
Fitting failure can occur for various reasons, such as a lack of available 
reads (due to later reads saturating) or fits that converged to a local, 
rather than a global, minimum. For both cases, we replace their coefficients 
with a 3-$\sigma$ clipped median of the correction coefficients in their 
individual detector quadrants.
\bigbreak
Finally, the data is reformatted to ensure compatibility with \texttt{calwf3}
execution. The structure of the flat reference file is described in 
Appendix~\ref{app:flt_dat_ref}. For technical details on the detector 
non-linearity correction applied during the \texttt{NLINCORR} step, refer 
to the 
\hyperlink{https://wfc3tools.readthedocs.io/en/latest/wfc3tools/calwf3.html}
{\texttt{calwf3} documentation}.
\begin{figure}[h!]
    \centering
    \includegraphics[width=1.0\linewidth]{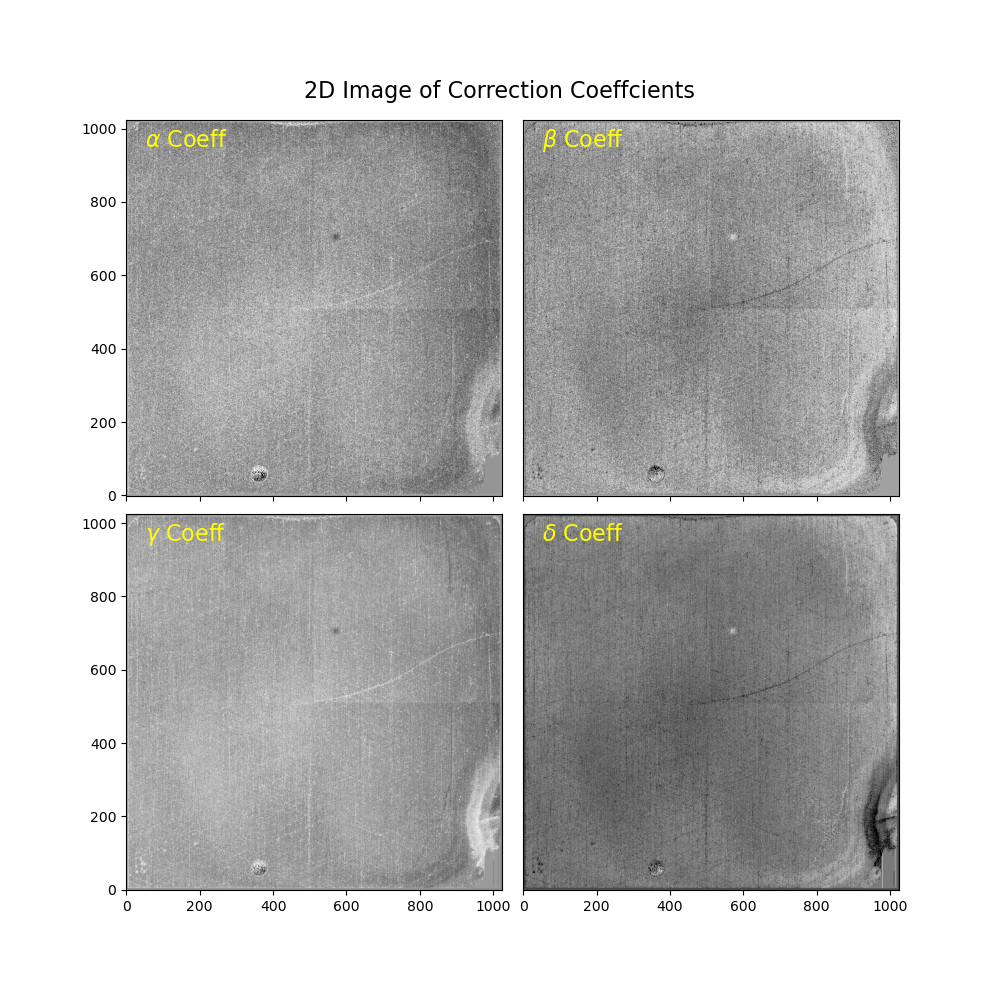}
    \vspace*{-0.8in}
    \caption{\small Maps of the four fit coefficients derived using the 
    model in Equation~\ref{eq:polyfit}. The coefficients are labeled at 
    the top of each image and highlight both pixel-to-pixel and large-scale 
    variations across the detector. In each image, the data is displayed 
    with a linear scale and zscale interval.}
    \label{fig:coeff}
\end{figure}
\begin{figure}[t!]
    \centering
    \includegraphics[width=1.0\linewidth]{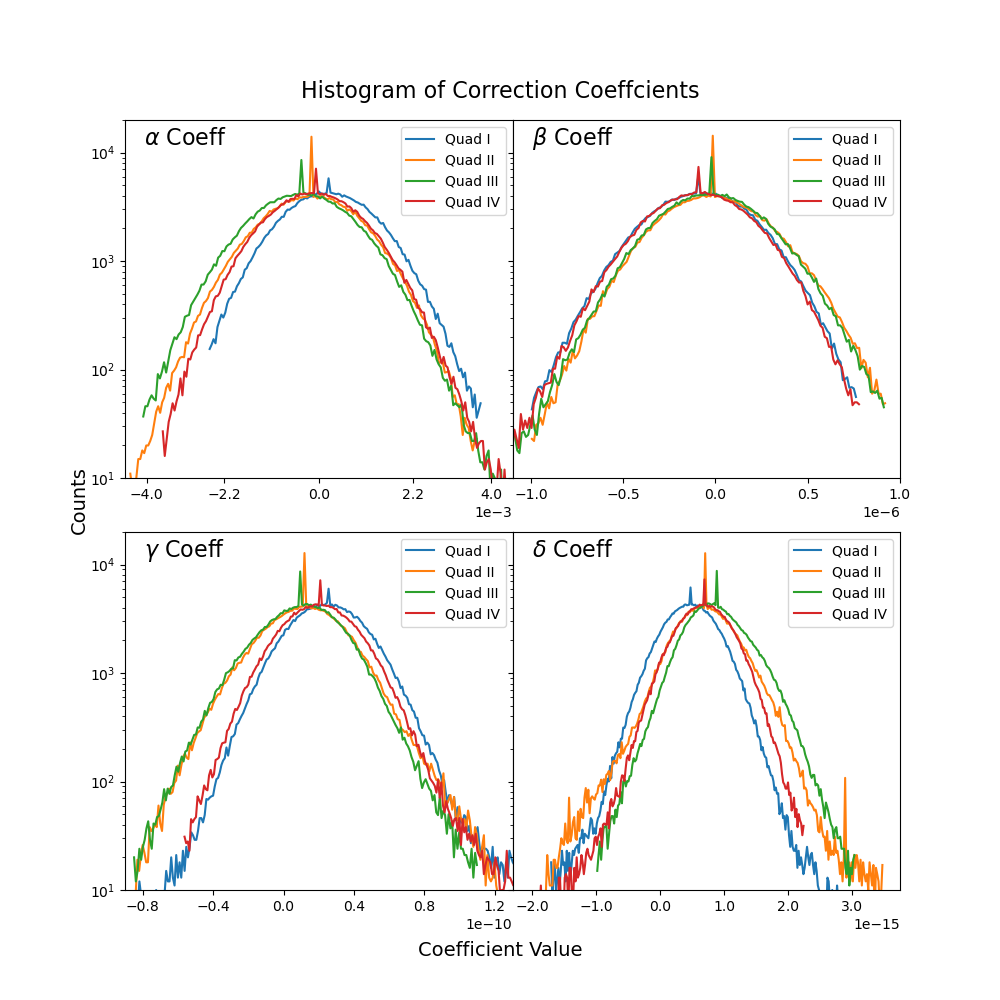}
    \vspace*{-0.3in}
    \caption{\small Histogram of the four non-linearity correction coefficient 
    values derived for each quadrant of the WFC3 IR detector using the model in 
    Equation \ref{eq:polyfit}.The four coefficients are labeled in the top-left 
    of each plot, and the quadrant IDs are labeled in the top-right box. The 
    spikes observed at the peak of each histogram originate from bad pixels 
    where the NaN values have been replaced with a quadrant-based average of 
    the correction coefficients.}
    \label{fig:coeff_hist}
\end{figure}
\bigbreak
Figure~\ref{fig:coeff} displays the four non-linearity correction 
coefficients from the pixel-based reference file. Because these coefficients 
were derived for individual pixels of the WFC3/IR detector, they exhibit 
the same detector artifacts observed in the raw and processed data (e.g., 
the wagon wheel, death star, and un-bonded pixels). Figure~\ref{fig:coeff_hist} 
shows histogram plots of these four coefficients. A notable feature in these 
plots is the spikes (high bin counts) overlaid on the peak of the histogram. 
These spikes result from replacing the NaN values for masked and poor-fit pixels 
with the quadrant-based averages of the correction coefficients, as described 
previously.

\section*{Analysis}
Detailed testing of the new pixel-based, non-linearity correction file is 
described in an accompanying report WFC3 ISR 2025-09 by \textcite{huynh2025}, 
available from the 
\href{https://www.stsci.edu/hst/instrumentation/wfc3/documentation/instrument-science-reports-isrs?itemsPerPage=1000}{WFC3 Instrument Science Reports page}. 
\bigbreak
Testing was performed for a variety of targets and observing modes, including: 
internal flat fields, star clusters, and HST flux standards (imaging and grism 
data) acquired in a wide range of sample sequences. In general, the new 
correction improves the linearity for sources with fluence levels above 
$\sim$50,000 $e^-$ ($\sim$20,000 DN), with improvements up to 7\% for pixels 
with fluence levels approaching the saturation limit of $\sim$80,000 $e^-$ in 
the last unsaturated reads of the IMA file.  The effect is smaller in FLT 
data products, with improvements of up to $\sim$0.5\% in a 3-pixel aperture 
radius. The pixel-based solution also significantly decreases the number of 
cosmic rays flagged during \texttt{calwf3}'s up-the-ramp fit, leading to 
improved photometric accuracy and higher signal-to-noise in the calibrated 
FLT data products. 
\bigbreak
The new correction has a negligible impact 
($\sim$0.1--0.2\%) on the published IR zeropoints by \textcite{Calamida2024} 
and will therefore be implemented in the $\texttt{calwf3}$ pipeline without 
updating the WFC3/IR image photometry reference file (IMPHTTAB). For 
additional details, see the ``IR Zeropoints" section of WFC3 ISR 2025-09 by 
\textcite{huynh2025}.

\section*{Caveats}
\label{sec:caveats}
Whilst this analysis yields a new non-linearity solution with significant 
improvement in the high-fluence regime, a number of caveats may still 
persist, and could be improved upon in future work.
\bigbreak
First, the limited number of samples obtainable in a single WFC3/IR 
exposure (15) limits the overall precision of the non-linearity 
calibration.  As a pixel approaches saturation, the non-linearity of the 
pixel's response to incident photons increases.   If a pixel reaches 
saturation between read$_i$ and read$_{i+1}$, the signal measured at 
read$_{i+1}$ cannot be used to quantify the non-linear response.  Thus, 
to measure the pixel's response right up to the saturation limit, a read 
must occur just before the signal reaches the saturation limit.  With 
full-well depths in the range of 70,000 - 80,000 $e^-$, and only 15 
samples (which can only occur evenly spaced at the end of the ramp), this 
is unlikely.  Thus, the high-fluence regime will have higher uncertainties 
than the mid-fluence regime.
\bigbreak
In addition, the relation between cosmic ray flagging and non-linearity 
is somewhat circular, as the cosmic ray flagging algorithm assumes the 
signal in a pixel accumulates linearly over the course of the ramp.  
However, if the signal is non-linear (even after applying the 
linearity calibration), any deviation from linearity can result in 
pixels being flagged in error as cosmic rays, even if the pixel did not 
receive a real cosmic ray impact.  As the creation of the non-linearity 
reference file requires cosmic rays to be masked (real jump discontinuities 
in the signal would invalidate the polynomial fits), cosmic ray hit 
pixels are removed before the creation of the master flat.  As a result, 
some of the most non-linear pixels may be systematically removed from this 
analysis (and thus not fully calibrated) due to erroneous cosmic ray 
flagging.  To fix this, an iterative approach would need to be applied, 
solving for the linearity coefficients and incrementally increasing the 
cosmic ray detection sensitivity repeatedly until a convergent solution 
is found.
\bigbreak
Lastly, the non-linearity in the low fluence regime adds further 
uncertainty to this calibration.  As this analysis assumes the first few 
reads are linear (with a potential nonzero intercept), the $y_{ideal}$  
signal used to normalize $y_{obs}$ introduces a small error to the 
correction in the low fluence end (1-2\%), where the correction is 
typically dominated by the linear term of the solution.  As the first few 
reads of the SPARS25 internal flats span the first 10-20\% of the full 
well depth, this error may affect the calibration for pixels imaging 
fainter sources. In theory, leveraging data taken using other sample 
sequences may allow for finer sampling in the low-fluence regime and allow 
for a reduction in this error.  However, combining data from the various 
sample sequences requires more careful consideration and is out of the 
scope of this document.

\section*{Summary}
\label{sec:summary}
We outline the methodology for creating a new pixel-based non-linearity 
correction for the WFC3/IR detector, which improves upon the current 
quadrant-based correction derived from ground test data. Utilizing 
SPARS25 internal flats acquired on-orbit over 3 years of CAL/WFC3 
non-linearity monitoring programs, we generate a bias-subtracted master 
flat.
\bigbreak
Using this master flat, we derive linearity correction coefficients for 
each pixel modeled by a third-order polynomial. We are unable to derive 
correction coefficients for $\sim$2\% of detector pixels which are known 
to be bad. For these, we replaced their values with quadrant-based averages 
calculated from the newly derived corrections.
\bigbreak
Testing by \textcite{huynh2025} shows that the current quadrant-based 
correction and the new pixel-based correction perform similarly for low 
fluence levels ($<$50,000 $e^-$). However, at higher fluence levels 
($\sim$50,000 - 80,000 $e^-$), the pixel-based correction improves the 
linearity by up to 7\% in the last reads of the IMA file just before 
saturation. The effect is smaller in FLT images after the ramp fit in 
$\texttt{calwf3}$, with improvements of up to $\sim$0.5\% in the 
observed photometry of bright stars in a 3-pixel aperture radius.
\bigbreak
The updated non-linearity reference file \texttt{9au15283i\_lin.fits} was 
delivered to the Calibration Database Reference System in October 2025 and 
will be used to reprocess all WFC3/IR imaging and grism observations in the 
MAST archive by late-2025.

\section*{Acknowledgements}
We appreciate the careful review of this work by Benjamin Kuhn, Joel 
Green, Norman Grogin, \& Sylvia Baggett and would like to thank them 
for their suggestions to improve the quality of this report. We would 
also like to thank the WFC3 Photometry team and Peter McCullough for 
help in critiquing this work and their helpful suggestions in improving 
this project.

\printbibliography

\begin{appendices}

\section{Input Manifest}
\label{app:in_mani}
The table below lists the input data that was used to generate the master 
flat used in deriving the non-linearity correction coefficients. All of 
the data was collected as part of three WFC3/IR Linearity Monitoring 
programs. The table only includes the identities of flats in F127M and dark 
observations; all of the other observations in each program are excluded 
from this work. Observations were taken from March 2011 to April 2013. 
\begin{longtable}{| c | l | c | c | p{6.5cm} |}
\caption{Flat field \& dark current exposures used 
to derive the non-linearity coefficients for the 
WFC3/IR detector.}
\label{tab:dataidflat} \\[3pt]
\hline
\hline
\bf{Proposal} & \bf{Filter}  & \bf{Sample} & \bf{Exposure} & 
\bf{Observation ID} \\[3pt]
\bf {ID} & & \bf{Sequence} & \bf{Time (sec)} & \\
\hline
\hline
\endfirsthead
\caption*{Table~\ref{tab:dataidflat} continued.} \\[3pt]
\hline
\hline
\bf{Proposal} & \bf{Filter}  & \bf{Sample} & \bf{Exposure} & 
\bf{Observation ID} \\[3pt]
\bf {ID} & & \bf{Sequence} & \bf{Time (sec)} & \\
\hline
\hline
\endhead
12352 & Blank & \shortstack{SPARS25,\\\\NSAMP=12} & 252.9 &
\shortstack{\\[3pt]
    ibmg02sfq, ibmg03a7q, ibmg04ceq, \\ 
    ibmg05dcq, ibmg06eaq, ibmg07jkq, \\ 
    ibmg10uvq, ibmg13gsq, ibmg14q3q, \\ 
    ibmg15qnq, ibmg16r9q 
    } \\[3pt]
\hline
12352  & F127M & \shortstack{SPARS25,\\\\NSAMP=16} & 352.9 &
\shortstack{\\[3pt]
    ibmg02siq, ibmg03aaq, ibmg04chq, \\
    ibmg07jtq, ibmg05dfq, ibmg06edq, \\
    ibmg14q6q, ibmg10uzq, ibmg13gvq, \\
    ibmg15qqq, ibmg16rcq
} \\[3pt]
\hline
12696  & Blank & \shortstack{SPARS25,\\\\NSAMP=12} & 252.9 & 
\shortstack{\\[3pt]
    ibvl01adq, ibvl02dqq, ibvl03f2q, \\
    ibvl04kbq, ibvl05o8q, ibvl13soq, \\
    ibvl14ovq, ibvl15fxq, ibvl16k2q
} \\[3pt]
\hline
12696  & F127M & \shortstack{SPARS25,\\\\NSAMP=16} & 352.9 & 
\shortstack{\\[3pt]
    ibvl01ahq, ibvl02dtq, ibvl03f5q, \\
    ibvl04khq, ibvl05obq, ibvl13suq, \\
    ibvl14p2q, ibvl15g0q, ibvl16k6q
} \\[3pt]
\hline
13079 & Blank & \shortstack{\\SPARS25,\\NSAMP=12} & 252.9 & 
\shortstack{\\[3pt]
    ic5n07jlq, ic5n08khq, ic5n09mwq
} \\[3pt]
\hline
13079 & F127M & \shortstack{\\SPARS25,\\NSAMP=16} & 352.9 &
\shortstack{\\[3pt]
    ic5n07joq, ic5n08kkq, ic5n09mzq
} \\[3pt]
\hline
\end{longtable}

\vspace{0.1in}

\section{Reference File Data Format}
\label{app:flt_dat_ref}
\begin{figure}[h!]
    \centering
    \includegraphics[width=.85\linewidth]{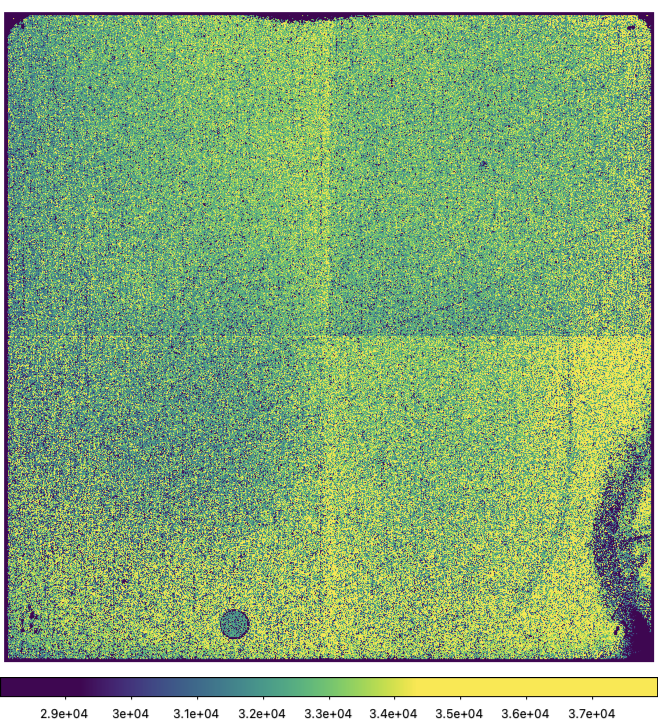}
    \caption{Saturation map defining the full well limit for every 
    pixel on the WFC3/IR channel. The color bar is in units of DN. 
    The saturation map is populated in the NODE extension (EXT = 16) 
    of the NLINFILE. The values in this extension are unchanged from 
    the original analysis by 
(\cite{hilbert2008}).}
    \label{fig:f_satmap}
\end{figure}

The non-linearity correction reference file, as described in 
\S\href{https://newcdbs.stsci.edu/doc/Section11.html#linearity-correction-file-lin-uniquename-lin-fits}
{11.2.8} of HST Reference Files Information Document, is stored in a 
standard FITS format with multiple COEF and ERR extensions and a single DQ, 
NODE, ZSCI, \& ZERR extensions with the same dimensions as the WFC3 IR 
detector array (1024x1024).
\bigbreak
\vspace*{-0.2in}
\begin{table}[h]
\caption{Linearity correction reference file data format.}
\label{tab:data_format}
\renewcommand{\arraystretch}{1.2}
\begin{tabular}{|c|l|c|c|c|l|}
\hline
Ext. No. & EXTNAME & EXTVER & BITPIX & Dimension & Description \\
\hline
\hline
0 & N/A & N/A & 8 & N/A & Primary header (no data) \\ 
1 & COEF & 1 & -32 & 1024x1024 & Linearity coefficient $\alpha$ \\ 
2 & COEF & 2 & -32 & 1024x1024 & Linearity $\beta$ \\ 
3 & COEF & 3 & -32 & 1024x1024 & Linearity $\gamma$ \\ 
4 & COEF & 4 & -32 & 1024x1024 & Linearity $\delta$ \\ 
5 & ERR & 1 & -32 & 1024x1024 & Variance $\alpha$ \\ 
6 & ERR & 2 & -32 & 1024x1024 & Variance $\beta$ \\ 
7 & ERR & 3 & -32 & 1024x1024 & Variance $\gamma$ \\ 
8 & ERR & 4 & -32 & 1024x1024 & Variance $\delta$ \\ 
9 & ERR & 5 & -32 & 1024x1024 & Covariance $\alpha\beta$ \\ 
10 & ERR & 6 & -32 & 1024x1024 & Covariance $\beta\gamma$ \\ 
11 & ERR & 7 & -32 & 1024x1024 & Covariance $\gamma\delta$ \\ 
12 & ERR & 8 & -32 & 1024x1024 & Covariance $\alpha\gamma$ \\ 
13 & ERR & 9 & -32 & 1024x1024 & Covariance $\beta\delta$ \\ 
14 & ERR & 10 & -32 & 1024x1024 & Covariance $\alpha\delta$ \\ 
15 & DQ & 1 & 16 & 1024x1024 & Data quality flags \\ 
16 & NODE & 1 & -64 & 1024x1024 & Saturation threshold \\ 
17 & ZSCI & 1 & -32 & 1024x1024 & Super zero read \\ 
18 & ZERR & 1 & -32 & 1024x1024 & Uncertainties in zero read \\
\hline
\end{tabular}
\end{table}
\bigbreak
The correction coefficients are stored in floating-point image extensions 
with EXTNAME set to COEF and EXTVER ranging from 1 to 4, denoting the 4 
coefficients from Equation ~\ref{eq:polyfit}.
\bigbreak
The polynomial model (Equation ~\ref{eq:polyfit}) fit variance of these 
coefficients are stored in the image extension with EXTNAME set to ERR 
and EXTVER ranging  from 1 to 4, while the covariance terms are saved with 
EXTNAME set to ERR and EXTVER ranging from 5 to 10. 
\bigbreak
The data quality flags are stored in [DQ, 1] extension of the FITS file. 
The saturation map, which defines the full well limit for each individual 
pixel, is contained in the NODE extension of the FITS file. 
Figure~\ref{fig:f_satmap} displays the map in units of DN and is unchanged 
from the analysis by (\cite{hilbert2008}) using saturated STEP50 flats 
acquired during ground testing. (The pixel values in this extension are 
identical in the quad-based and in the pixel-based NLINFILE).
\bigbreak
The last two extensions of the non-linearity correction reference file 
contains the super zero read and its statistical error images with EXTNAME 
ZSCI \& ZERR and EXTVER set to 1. For ease of comprehension, the structure 
of the non-linearity correction reference file is also shown in a table 
format in Table~\ref{tab:data_format}.

\end{appendices}
\end{document}